\newcommand{\beq}{\begin{equation}}
\newcommand{\eeq}{\end{equation}}
\newcommand{\pa}{p^{(0)}}
\newcommand{\pb}{p^{(1)}}
\newcommand{\etal}{{\it et. al.}}
\newcommand{\Ninput}{N_d}
\newcommand{\omid}{o_{\rm mid}}
\newcommand{\FC}{{f_C}}
\newcommand{\tv}{{\tt t}}
\newcommand{\atv}{\bar{\tt t}}
\newcommand{\kk}{{\,k}}

\newcommand{\Prop}{{\cal P}}

\def\wisk#1{\ifmmode{#1}\else{$#1$}\fi}
\def\deg    {\wisk{^\circ}}

\documentclass[preprint]{aastex}

\begin{document}

\title{Neural networks as a tool for parameter estimation
       in astrophysical data}

\author{Nicholas G. Phillips
	\footnote{email: {\tt Nicholas.G.Phillips.1@gsfc.nasa.gov}}}
\affil{SSAI, Laboratory for Astronomy and Solar Physics, 
	Code 685, NASA/GSFC, Greenbelt, Maryland 20771}
\and
\author{Alan J. Kogut
	\footnote{emai: {\tt Alan.J.Kogut.1@gsfc.nasa.gov}}}
\affil{Laboratory for Astronomy and Solar Physics, 
	Code 685, NASA/GSFC, Greenbelt, Maryland 20771}

\date{\today}

\begin{abstract}
We present a neural net algorithm for parameter estimation
in the context of large cosmological data sets.
Cosmological data sets present a particular challenge
to pattern-recognition algorithms
since the input patterns
(galaxy redshift surveys,
maps of cosmic microwave background anisotropy)
are not fixed templates overlaid with random noise,
but rather are random realizations
whose information content lies 
in the correlations between data points.
We train a ``committee'' of neural nets
to distinguish between Monte Carlo simulations
at fixed parameter values.
Sampling the trained networks
using additional Monte Carlo simulations
generated at intermediate parameter values
allows accurate interpolation
to parameter values for which the networks were never trained.
The Monte Carlo samples automatically provide
the probability distributions and truth tables
required for either a frequentist or Bayseian analysis
of the one observable sky.
We demonstrate that neural networks
provide unbiased parameter estimation
with comparable precision as maximum-likelihood algorithms
but significant computational savings.
In the context of CMB anisotropies, 
the computational cost for parameter
estimation via neural networks scales as $N^{3/2}$.
The results are insensitive to the noise levels
and sampling schemes typical of large cosmological data sets
and provide a desirable tool for the 
new generation of large, complex data sets.
\end{abstract}
%===================================================================%
%                     Section: Introduction                         %
%===================================================================%
\section{Introduction}

%---------- role of parameter estimation in astrophysics ------------

A fundamental question in cosmology is the origin and evolution  of large scale
structure in the universe.   The standard model for this evolution is the
gravitational growth and collapse  of initially small perturbations in the
primordial density distribution. This picture is supported by the detection of 
primordial Cosmic Microwave Background (CMB)
temperature anisotropies at a level of approximately one part in
$10^5$ by the Cosmic Background Explorer (COBE) satellite and a series of ground-based
and balloon-borne experiments. 
Small perturbations on the matter and energy density in the early universe are
reflected in the temperature distribution of the CMB, providing a ``snapshot''
of conditions the early universe while the perturbations were still in the
linear regime.

% Angular scales

One angular scale of particular interest is the horizon size at the  surface of
last scattering, the epoch when the universe cooled sufficiently to form
neutral hydrogen and allow the CMB photons to propagate freely.
Causally-connected regions at the surface of last scattering, as viewed from
the present epoch, subtend an angle
$$
\theta ~\sim ~1.7\deg ~\Omega_0^{1/2} ~( \frac{1100}{1 + z_{ls}} )^{1/2}
$$
\noindent
where $z_{ls}$ is the redshift at last scattering and $\Omega_0$ is the total
density of the universe relative to the critical (closure) density. Anisotropy
on scales larger than $\sim 2\deg$ ~reflect perturbations larger  than the
particle horizon and thus probe the primordial density distribution. On scales
smaller than 2\deg, causal mechanisms become important and modify the
primordial density in model-specific ways.

%---------- review of max(L) method ------------
% Want to estimate parameters

From the one observable sky, we want to infer the values of these
cosmological parameters with minimal uncertainty in the shortest possible time.
Theoretical models do not predict a specific template for the CMB anisotropy (a
hot spot at this location, a cold spot over there), but rather predict a
statistical distribution usually expressed in terms of the angular power
spectrum. Deriving the power spectrum from the data (or more generally,
deriving model parameters directly from the sky maps) involves accounting for
angular correlations between pixels, precluding use of simple linear
least-squares techniques. The accepted standard in the CMB community has been
the generalization of least-squares techniques as implemented in maximum
likelihood algorithms (see, e.g.,
G\'{o}rski et al.\ 1996;
Tegmark, Taylor, \& Heavens 1997;
Bond, Jaffe, \& Knox 1998;
Borrill 1999).

% Simplest method is least squares

The simplest method of parameter estimation uses a
goodness-of-fit test to compare a set of observables  $y_i \pm \sigma_i$
measured at a set of positions $x_i$ to a theoretical model $\Gamma_i$. If we
have $N_d$ data points $y_i$ and $N_p$ parameters $p_j$, we define
\begin{equation}
\chi^2 = \sum_{i=1}^{N_d}
\left( 
\frac{ y_i - \Gamma_i }{\sigma_i} 
\right)^2,
\label{simple_chi}
\end{equation}
where 
\begin{equation}
\Gamma_i(x) = \sum_{j=1}^{N_p} p_j X_j(x_i)
\label{linear_eq}
\end{equation}
is function of the parameters $p$
and some fixed basis functions $X(x)$.
We obtain the ``best-fit'' parameter values
by minimizing $\chi^2$ with respect to the parameters,
\begin{equation}
\frac{\partial \chi^2}{\partial p_j} = 0
\label{min_chi_eq}
\end{equation}
for the $j^{\rm th}$ parameter $p_j$. The least-squares system in Eq.
\ref{min_chi_eq} has the solution
\begin{equation}
p_j = \sum_{k=1}^{N_p} ~({\bf A}^{-1})_{jk} B_j
\label{normal_eq}
\end{equation}
where
\begin{equation}
{\bf A}_{jk} = \sum_{i=1}^{N_d} 
~ \frac{ X_j(x_i) X_k(x_i) }{ \sigma_i^2 }
\end{equation}
is an $N_p \times N_p$ matrix, and
\begin{equation}
B_k = \sum_{i=1}^{N_d} 
~ \frac{ y_i X_k(x_i) }{ \sigma_i^2 }
\end{equation}
is a vector of length $N_p$. 

% Generalize to max likelihood

If the data points are not independent, this relatively simple calculation
becomes much more costly. Covariance between the observed data points can
result from instrumental artifacts (correlated noise, instrumental resolution,
oversampled data) or from correlations in the underlying signal (for instance,
measuring in real space  a signal whose components are independent in Fourier
space). Equation \ref{simple_chi} can be generalized  to include the effects of
covariance,
\begin{equation}
\chi^2 = \sum_{i=1}^{N_d} \sum_{j=1}^{N_d}
(y_i - \Gamma_i) ({\bf M}^{-1})_{ij} (y_j - \Gamma_j),
\label{chisq_def}
\end{equation}
where
\begin{equation}
{\bf M}_{ij} = \langle 
~(y_i - \langle \Gamma_i \rangle)
~(y_j - \langle \Gamma_j \rangle) ~\rangle
\label{covar_def}
\end{equation}
is the $N_d \times N_d$ covariance matrix  and the brackets denote an ensemble
average.

Conjugate gradient techniques can solve for $\chi^2$ without expliciting
solving for ${\bf M}^{-1}$ and thus avoiding the  $O(N_d^3)$ operations
this would incur. But if the covariance matrix ${\bf M}$ depends on the
parameters $p_j$, then minimizing $\chi^2$ will produce biased estimates
for $p_j$.
Maximum-likelihood parameter estimation provide a tool  to overcome 
this limitation.
For a multivariate Gaussian
distribution, the probability of obtaining the observed data $y_i$ given a set
of parameters $p_j$ is
\begin{equation}
{\cal L} = P(y | p) =
(2\pi)^{-N_d/2} ~
\frac{ \exp( -\frac{1}{2} \chi^2 )}
     { |{\bf M}|^{1/2} }
\label{like_def}
\end{equation}
where $\chi^2$ is defined in Eq. \ref{chisq_def}. The ``best'' choice of
parameters is that which maximizes the likelihood function ${\cal L}$. The
curvature of the likelihood surface about the maximum defines the uncertainty
in the fitted parameters,
\begin{equation}
\delta p_j \geq \sqrt{ ({\bf F}^{-1})_{jj} }
\label{err_def}
\end{equation}
where
\begin{equation}
{\bf F}_{ij} = \langle
\frac{ \partial^2 L }
     { \partial p_i \partial p_j } \rangle
\label{fisher_def}
\end{equation}
is the Fisher information matrix
(Kendall \& Stuart 1969)
and $L = -\log({\cal L})$
(see Bunn \& Sugiyama 1995;
Vogeley \& Szalay 1996;
Tegmark et al.\ 1997;
Bond et al.\ 1998).

The maximum likelihood estimator is unbiased and asymptotically approaches the
equality in Eqn. \ref{err_def}. However, these advantages come at a steep price:
both the $\chi^2$ and the determinant calculation  in Eq. \ref{like_def} scale
as $O(N_d^3)$, making brute-force techniques computationally infeasible. 
For $N_d > 10^6$ the time required  is measured in years, 
even on the most powerful supercomputers. 
A number of authors have suggested ways around this limitation,
see {\it e.g.} 
Oh Spergel \& Hinshaw 1999,
Hivon, \etal \ 2001 and
Dor\'{e}, Knox \& Peel 2001.
Such methods usually rely on symmetries in the data, {\it i..e}, axial
symmetry of the noise, while neural networks need to make no assumptions
about symmetries.

%---------- previous work -> template matching ------------

Neural networks provide an alternative for astrophysical parameter
estimation.They have been used previously in astronomy 
for galaxy classification \cite{Lahav96,Andreon00} and
periodicity analysis of unevenly sampled data as applied to
stellar light curves \cite{Tagliaferri00}.
They have also been used to analyze
stellar spectra \cite{BailerJones97,BailerJones98,BailerJones00},
with results comparable to traditional methods. 
However, Bailer-Jones {\it et. al.}
compared data to a deterministic model (stellar spectra),
whereas cosmological applications examine random patterns
drawn from parameterized stochastic models.
We demonstrate the
generality of our algorithm by considering different problems with
the same network architecture.
We find the computational cost for training the network,
in the context of CMB anisotropy,
requires $O(N^{3/2})$ operations and thus provides a substantial improvement
over brute-force maximum-likelihood methods.

%   stochastic nature

The stochastic nature of our models is fundamental: their starting
point is the quantum nature of the early universe. What the models predict
are the parent populations from which individual realizations are to be drawn.
Our single observed universe is assumed to be such a realization. The
question becomes: given a particular model, which parameters that define the
parent population best describe the observed data?
This is to be contrasted with the
problems of steller spectra or galaxy classification, which have at their heart
either a deterministic model or a pre-assigned catalogue of types.

%   not always obvious what is best statistic: topology, non-Gaussian physics

This stochastic nature means any analysis of the observed data must rely
on some statistical test. Leaving aside for the moment the computational
challenges of traditional maximum likelihood methods, this raises the 
problem of knowing the best statistic for distinguishing between different
parameters or models (we can view model selection as a discrete parameter to be 
estimated). Maximum likelihood methods require an {\it a priori} definition 
of  a goodness-of-fit function.
The choice of a goodness-of-fit function is not always  obvious and
is particularly acute for 2D and 3D surveys.
Much of the information lies in the {\it phase} features of these surveys.
Statistical tests can fail badly in detecting phase features, as witness the
large literature devoted to the relatively simple problem of edge detection in
2D data sets (see, e.g., Hough\ 1962 and Davis\ 1975).
Topological tests
such as the genus or other Minkowski functionals have been applied to 2- and
3-D maps, \cite{Gott90,Kogut96} but the relative power of these statistics is poor
\cite{PhillipsKogut01}. Neural networks, in contrast, do not require specification of
a single statistic of {\it a priori} interest.
As the network is trained, it determines how it will discriminate
between competing models.

% naturally include instrumental effects without problems

All observations add instrumental effects to the desired physical signal.
Any analysis of the data must model these effects. For neural networks, this
present no great challenge as long as we can model the effects. For either 
irregularly sampled time series data or 2D/3D survey data with missing patches,
any analysis based on Fourier methods will suffer from alaising of power.
With neural networks, this problem does not arise; by excluding from the
simulations the data points missing in the observed data, the networks will
learn the effects of the gaps. The inclusion of noise to the simulations
may increase the number of training passes needed, but will not prevent
the use of neural networks for parameter estimation.

%===================================================================%
%                     Section: Neural Network                       %
%===================================================================%
\section{Neural Network}
Neural networks can be trained to estimate the value of a continuous parameter,
and can reliably interpolate to parameter values intermediate between the 
training values.  Though this idea is not new to astrophysics, {\it e.g.}
\cite{BailerJones97,BailerJones98,BailerJones00}, our method is fundamentally
different from that of Bailer-Jones {\it et. al.} in that our patterns are
random samples drawn from a parameterized parent population. The Bailer-Jones
{\it et. al.} method is a matter of template matching; randomness only enters
their input patterns as instrument noise. 
For CMB and other cosmological data, the patterns themselves are
intrinsically random. Nonetheless, using the same basic
neural network architecture, we can train the networks to discriminate
stochastic patterns that differ according to the parent population from which
they are drawn.

We start by training a network to differentiate between  simulated data sets
(including instrument noise and other artifacts) generated at a pair of
discrete parameter values. The back-propagation adjusts the weights until the
network outputs target value 0 when presented with the first set of patterns,
and target value 1 when presented with the second. In practice, since the
information distinguishing different parameter values is in the correlations,
not the actual pixel values, any single network will not train to a sufficient
degree. We improve the situation by training a small committee of networks and
polling them to get a consensus opinion. Now we find simulations generated
at the training parameter values produce two well defined peaks.
Simulations generated with an
intermediate parameter value (never present in training data) yield outputs
peaking somewhere in between, depending on whether the new parameter value is
closer to the first or second  training value (see Fig \ref{sample-outputs}).

We quantify this behavior by presenting the trained network with a set of new
inputs drawn from a grid of intermediate parameter values, and derive for each
intermediate parameter value the corresponding probability distribution of
output values. When the networks are later presented with an  unknown pattern,
each distribution gives the probability that the unknown  pattern was generated
with a parameter value corresponding to that grid point. The interpolated
parameter is the probability-weighted mean. The grid samples  all use the same
trained networks; the sampling of the networks at different  grid points is
faster than the training since we usually need many more training sets than
sampling sets, thus no great  computational cost is incurred. 
This sort of sampling of trained
networks becomes a key component in utilizing a Bayesian approach to
parameter estimation, see Christensen and Meyer\ 2000 and Rocha \etal \ 2000
for discussions of
the Bayesian approach in the context CMB anisotropies and
MacKay 1995, along with  Bishop\ 1995 for neural networks.
Although we focus below on estimating a single parameter, 
the method is readily extended to multi-parameter fits.

%---------- SHORT statement of type of network: MLP ------------
\subsection{Parameter estimation via the Bayesian approach}

We use the standard MLP: $y_k^l = f(\sum_{i=0}^{N_{l-1}} w^l_{ki} y_i^{l-1})$,
where $y_k^l$ is the outout of neuron $k$ for layer $l$,
$N_l$ is the number of neurons in layer $l$ and
$w^l_{ki}$ is the weight connecting neuron $i$ of layer $l-1$ with
neuron $k$ of layer $l$. We include a bias for each neuron via $y_0^l = 1$. 
The activation function is the sigmoid $f(x) = 1/(1+e^{-x})$.
We only consider networks with the fully-connected, single hidden layer topology
and a single output neuron. The number of weights per network is
$N_{\rm wgt} = N_{\rm hidden}(N_{\rm input}+2)+1$.

We use standard backpropagation for training, with weight updates after each
training point. The quadratic cost function is used:
$E = \frac{1}{2}(o_{\rm patt} - t_{\rm patt})^2$, where 
$o_{\rm patt} = y^{\rm output\;layer}({\bf X}_{\rm patt})$ is the network
output when presented with the input ${\bf X}_{\rm patt}$ and $t_{\rm patt}$
is the desired target for that pattern.

%---------- Details for doing parameter estimation: ------------

The input patterns ${\bf X}$ are the end product of our simulations and as
such are realizations taken from an underlying parameterized probability 
distribution with parameter $p$. Thus we label each input pattern by the
parameter from which it was drawn: ${\bf X}(p)$.
Assuming the data lies
between two extreme values $\pa$ and $\pb$ we train a network to the target
$t^{(0)}=0$ for realizations ${\bf X}(\pa)$ and $t^{(1)}=1$ for ${\bf X}(\pb)$
by repeatedly presenting the network with new samples at $\pa$ and $\pb$ and
back propagating the error.  Once trained, the network will give an output
%$o\left({\bf X};\pa,\pb\right)$ 
$o\left({\bf X}\right)$ between 0 and 1 for any input parameter value. 
If $p < \pa$, the output clips at 0, while if $p > \pb$ the output clips at 1. 

\begin{figure}
\plotone{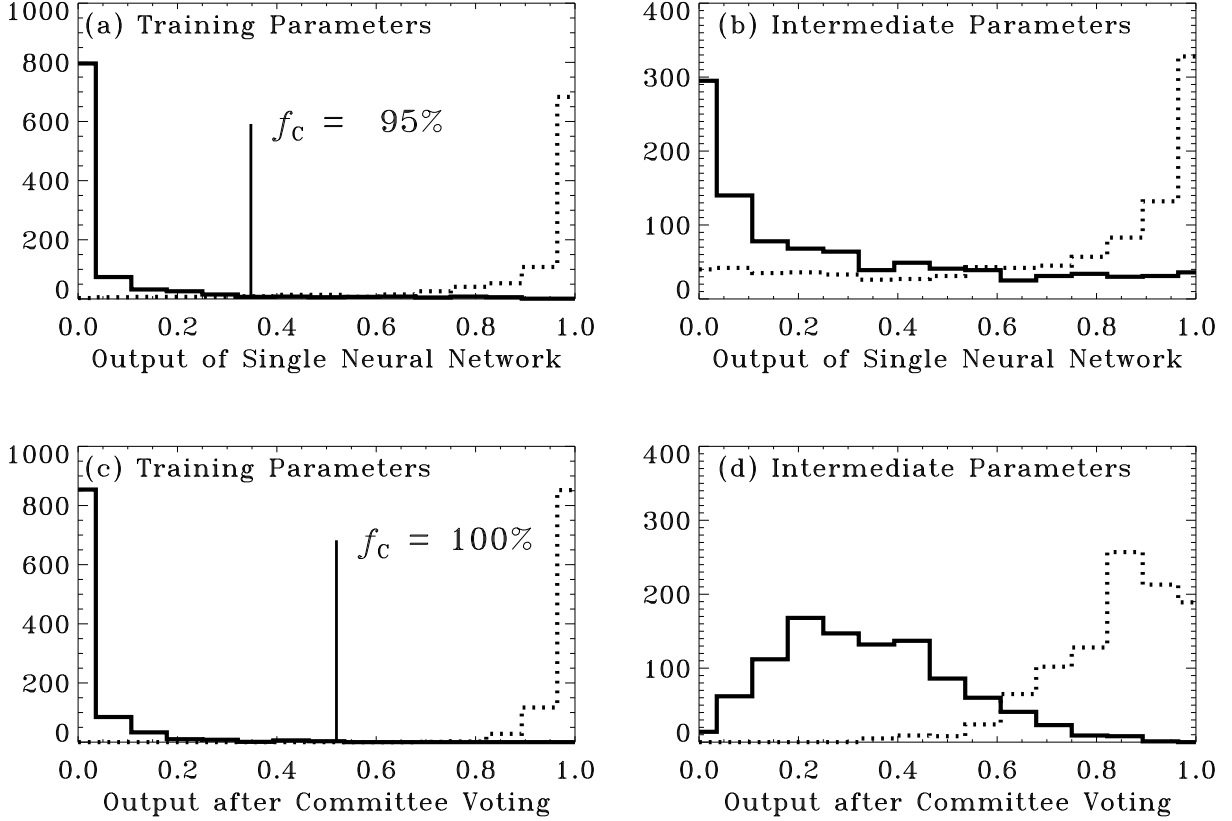}
\caption{
Sample neural network output distributions.
(a) Solid line is the distribution of output values for an independent set of
samples drawn at the same parameter value $\pa$ for which the network was
trained to the target $0$. The dotted line is for samples drawn
at $\pb$, the value for the target $1$. The vertical line is
the midpoint value $\omid$ that maximizes the fraction correct $\FC$, 
{\it i.e.}, all output values $<\omid$ are identified as being drawn at 
$\pa$ while the rest at $\pb$. 
(b) The output of the same network, but for samples drawn at parameters 
intermediate to $\pa$ and $\pb$. The solid line is for $p$ close to $\pa$ 
and the dotted line for a choice close to $\pb$. 
(c) Solid line is the average truth value $\atv$ for the $\pa$ patterns, 
averaged over a committee of 50 networks, the only difference between
the networks being the initial randomization of the weights.
The dotted line is for $\pb$.
(d) The average truth value for the same sets of samples in (b).
The averaging has produced two well defined 
peaks that are cleanly separated.
Distributions of $\atv$ like those in (c) and (d) 
become the basis for predicting which estimated parameters to 
associate with the average truth values, the ouput due to presenting a 
committe of networks with an unknown pattern.
}
\label{sample-outputs}
\end{figure}

Once we have trained a network, we can present additional, statistically
independent samples drawn at $\pa$ and $\pb$. Figure \ref{sample-outputs}a
shows the output distributions for $1000$ patterns of each parameter. 
We see the network has successfully trained in that the $\pa$ distribution
is peaked at $o=0$ while the $\pb$ samples at $o=1$.
To be able to interpolate to intermediate values, we will need to present
samples drawn at intermediate parameter values. Fig \ref{sample-outputs}b
shows the output distributions for an additional $1000$ samples each for two
parameters, one just a little larger than $\pa$ and one a little smaller than
$\pb$. These distribution also show the same tendency to peak close to the
limits of $0$ and $1$, but not as strongly as those drawn at the trained
parameters. In effect, the network is chosing which of the training parameters
these new patterns, for which it was never trained, most closely resemble.
At it stands now, this tendency makes it hard to construct the probability
distributions we need for parameter estimation.

By using a committee of networks, we take advantage of this peaking tendency.
We want to determine the committee consensus and from this get the distributions
we seek. The first step is converting the continuous output value into a discrete
{\it truth values} $0$ or $1$.
For each trained network,
we associate a midpoint value $\omid$ and for any input pattern ${\bf X}$,
we define its truth value according to 
\beq
\tv \equiv \tv\left({\bf X}\right) = \left\{
\begin{array}{cc}
  0 ; & o\left({\bf X}\right) \le \omid \\
  1 ; & o\left({\bf X}\right)  >  \omid
\end{array}
\right..
\eeq
We interpret $\tv({\bf X}) = 0$ as indicating the pattern was drawn from the parent
population with parameter $\pa$, and similarly we associate $\tv=1$ with $\pb$.

To determine $\omid$, we present $N$ samples drawn at $\pa$ and $N$ samples
at $\pb$. For each of these sets and any $\tilde\omid$, we obtain the
truth values $\tv_i^{(0)}$ and $\tv_i^{(1)}$, $i=1,\ldots,N$.
With this, ${\tt n}^{(0)} = \sum_i(1-\tv^{(0)}_i)$ is the number patterns drawn
at $\pa$ correctly identified as drawn at $\pa$ and 
${\tt n}^{(1)} = \sum_i \tv^{(1)}_i$ similarly at $\pb$. We chose $\omid$ to
maximize $\FC = \frac{1}{2N}({\tt n}^{(0)} + {\tt n}^{(1)})$ and refer
to $\FC$ as the {\it fraction correct}, our main measure of how well a network
has trained. In Fig \ref{sample-outputs}a, $\omid$ is marked with the vertical
line and we find $\FC=94\%$.

We note $\FC$ allows us to determine the optimal number
of training passes, $N_{\rm Train}$.
Starting with an initially randomized
network, as the network trains, we intermittently pause the training and
sample the network to determine $\FC$. It steadily increases to a maximum value 
and then levels out: the minimum for the training error $E$ has been reached. 
We take $N_{\rm Train}$ to be just where this plateau starts
and thus avoid overtraining.

For each network, any given input pattern is converted into discrete
truth values $\tv$. We now form a committee of such networks, where the
only difference between the networks is the initial randomization of the
weights. We find committee sizes $N_{\rm net}\sim 50$ sufficient. After
presenting any given pattern ${\bf X}$ to the committee, we have the
collection of truth values $\tv({\bf X})_m$, $m=1,\ldots,N_{\rm net}$.
We view each $\tv_m$ as the vote from network $m$ as to whether the
pattern resembled those drawn at $\pa$ or $\pb$. The committee consensus
is formed by generating the {\it average truth value}
\beq
\atv({\bf X}) = \frac{1}{N_{\rm net}}\sum_{m=1}^{N_{\rm net}}\; \tv({\bf X})_m.
\eeq
Figure \ref{sample-outputs}c shows the distribution of average truth values
for the same set of samples drawn at $\pa$ and $\pb$ used in
Fig \ref{sample-outputs}a. They are now even more sharply peaked about
$\atv=0$ and $\atv=1$ and in terms of the average truth value, $\FC=100\%$.
More important are the distributions displayed in Fig \ref{sample-outputs}d.
These are for the same intermediate samples as used in Fig \ref{sample-outputs}b;
we have now two well defined distributions with peaks intermediate to the
peaks for the $\pa$ and $\pb$ samples. This is the general trend 
when we work in terms of the average truth value:
as the parameter $p$ is swept
from $\pa$ to $\pb$, we get well defined distributions whose peak moves from
$\atv\sim 0$ to $\atv\sim 1$.

When we present our committee with an observed pattern, ${\bf X}_{\rm obs}$, for
which we do not know the parameter, we get its average truth value $\atv_{\rm obs}$.
From the Bayesian viewpoint, we are interested in the posterior
distribution $\Prop(p|\atv_{\rm obs})$: given the observation, what is the 
probability the true parameter is $p$. With this, it is
straightforward to determine the mean estimated parameter for this pattern,
along with our confidence for this estimate. We use Bayes Theorem to express
this in terms of the prior distributions,
\beq
\Prop(p|\atv_{\rm obs}) = 
           \Prop(\atv_{\rm obs}|p)\frac{\Prop(p)}{\Prop(\atv_{\rm obs})},
\label{eq-bayes-thm}
\eeq
all of which are readily determined by sampling our committee.

Selecting $K$ parameter values
uniformly distributed between the training values,
$p^\kk = \pa,\;\pa + \Delta p,\;\pa + 2\Delta p,\;\ldots,\pb$, we
generate $N$ samples at each of these parameter values.
The samples are presented to the committee of networks and thus
for each sample ${\bf X}_i$ drawn at each parameter value $p^\kk$, its
average truth value $\atv_i^k$ is computed. 
The $\atv$'s will take on discrete values,
$\atv = 0,1/N_{\rm net},2/N_{\rm net},\ldots,1$,
and we let $n^k_j$ be the number of
samples drawn at $p^\kk$ with $\atv^k_i = j/N_{\rm net}$.
Since we have the same number of patterns in each of the $K$ sample sets,
the prior for the parameter is essentially uniform:
$\Prop(p) = \sum_k \delta(p-p^\kk)/K$.
The probability of getting any of the possible discrete values of $\atv$ is
proportional to the total number of times it occurs for all the samples:
$\Prop(\atv_j) = \frac{1}{K N}\sum_k n^\kk_j$.
And finally, given the true input parameter value is $p^\kk$, the probability 
of the committee generating the average truth value $\atv_j$ is given by
$\Prop(\atv_j|p^\kk) = n^\kk_j/N$. Taken together, we have the
posterior probability distribution
\beq
\Prop(p|\atv_j) = 
   \frac{\sum_k \delta(p-p^\kk) n^\kk_j}
        {\sum_{k'} n^{k'}_j}.
\label{eq-prob}
\eeq
For our observed pattern, we have the parameter estimate
\beq
p_{\rm obs} = \frac{ \sum_k p^k\,n_j^k }{ \sum_k n_j^k },
\eeq
{\it i.e.}, the parameter-weighted mean. We also get 
the 68\% confidence width for our
estimate without any extra work:
\beq
\sigma^2(p_{\rm obs}) =  
	\frac{\sum_{\,k} (p^\kk)^2 \; n_{\rm obs}^\kk}
         {\sum_{\,k} n_{\rm obs}^\kk} 
   - (p_{\rm obs})^2
\eeq
Having the probability distribution Eqn \ref{eq-prob} allows us to
determine explicit upper and lower confidence intervals, along with
the standard error of our parameter estimate.

So far we have outlined how to make a single pass through our parameter
estimation method. We must also determine when we have converged on the
best estimate. There are three ways the results generated by the above
algorithm can be sub-optimal: 
i) the true parameter is is either too close to either end of the 
   training range $\left[\pa,\pb\right]$ or outside it; 
ii) the training range is too broad; or
iii) the training range is too narrow.
The distributions used to determine Eqn \ref{eq-prob} also allow
us to test for each of these cases. If the first case did occur, then
we have $\left|p^{(i)} - p_{\rm obs}\right| < \sigma(p_{\rm obs})$
for either $i=0$ or $i=1$.
Assuming this is not the case, if
$\sigma(p_{\rm obs})$ is much smaller than $\pb-\pa$,
then we have too broad a training range. We should chose a new training
range centered around the current estimate $p_{\rm obs}$ 
and a width comparable to $\sigma(p_{\rm obs})$.
A too narrow training range is indicated by 
$\sigma(p_{\rm obs})\sim \pb-\pa$ and a new broader range needs to be used.

We find it beneficial to supplement the above tests by presenting the
committee of trained networks with an independent set of sample patterns
drawn at the estimated parameter. The distribution of estimated parameters
for this set should not peak near either end of the training range and
have a width comparable to the training range. The range is too narrow
if the distribution does not have a well-defined shape; is flat around the
range. In practice convergence
typically requires only two or three iterations.

%===================================================================%
%                     Section: Application                          %
%===================================================================%
\section{Application: Cosmic Microwave Background Maps}

To illustrate this algorithm, we
fit the spectral index for CMB anisotropies based on a Gaussian model,
an example of the type of problem that will need accurate methods with low
computational cost. 
Deriving cosmological parameters from maps of the cosmic microwave background
usually involves maximum likelihood algorithm whose computational cost ($N^3$)
makes them prohibitive for mega-pixel maps. 
Our neural network method provides 
rapid parameter estimation for
CMB maps.

%---------- COBE-DMR-esque CMB maps ------------

On angular scales $\theta > 2\deg$, anisotropy in the cosmic microwave
background   corresponds to primordial density perturbations with scale-free
power spectrum  $P_k \propto k^n$, where $k$ is the wavenumber  and $n$ is the
power-law index. This comes about as quantum fluctuations of the metric during
inflation are pushed outside the causal horizon, becoming classical in the
process. Later, after inflation has ended, the causal horizon catches up to
the fluctuations and they reenter.

We expand full-sky anisotropy maps in terms of spherical harmonics:
\beq
\frac{\Delta T}{T}(\theta,\phi) = \sum_{l,m} a_{lm} Y_{lm}(\theta,\phi).
\label{eq-spher}
\eeq
Since the primordial fluctuations are assumed to have amplitudes that are
independent Gaussian processes for each wavenumber, maps are readily generated
by projecting these processes onto the expansion coefficients $a_{lm}$. This
amounts to letting the $a_{lm}$'s be random Gaussian variables with
zero mean and the variance \cite{Bond87}
\begin{equation}
\left<|a_{lm}|^2\right> = 
           \left(\frac{ 5 Q^2 }{4\pi}\right)
           \frac{ \Gamma(l + (n-1)/2) \Gamma( (9-n)/2) }
                { \Gamma(l + (5-n)/2) \Gamma( (3+n)/2) }.
\label{eq-HZ}
\end{equation}
Our maps are parameterized by the spectral index $n$ and the amplitude $Q$.
We now see clearly why our maps are stochastic patterns: it is
only the statistical properties that are specified by our model. 
Moreover, these properties are not local for it is in the harmonic
conjugate space of our maps that we draw independent random variables.
Our networks will learn these non-local statistical correlations
as they are trained on different realizations from the above
parameterized distributions.

%---------- Instrument modeling: ------------
\subsection{Instrument modeling}

To be able to estimate the parameters for observed data sets, we need to
include instrument effects to the pure theory sky maps we generate. The
first effect is the pixelization of the full sky. We use the pixelization
scheme used by the {\it COBE}-DMR data, which results in a total of 6144 pixels, 
each of size $2.5^\deg\!\!\times\!2.5\deg$. To generate a single realization,
to each $a_{lm}$, up at an $l_{max} = 20$, we assign a Gaussian random number with
width given by Eqn. \ref{eq-HZ} and evaluate Eqn. \ref{eq-spher} with the
angular coordinates centered on each pixel: 
$(\theta,\phi) \rightarrow (\theta_i,\phi_i)$, $i=1,\ldots,6144$.
This pixel resolution oversamples the resolution of the radiometer horn.
To account for the radiometer horn profile,
we smooth each of these full-sky maps with a
Gaussian profile with a FWHM of $7^\deg$.

As true for all measurements, the detectors introduce noise into the signal.
The scanning strategy for COBE was to visit each pixel many times. Each visit
is accompanied by the introduction of Gaussian noise to the signal. Since
each addition of noise to each pixel is an independent random process, the
more often a pixel is visited, the more the noise is supressed. We model this
by adding  Gaussian noise to each pixel with the variance $(20~ {\rm
mK})^2/N_{{\rm obs},i}$,
where is $N_{{\rm obs},i}$ is the number of observations of pixel $i$
\cite{Bennett96}. (In truth, the pixels are visited in pairs and differences
measured; the analysis of the above reference takes this into account.)

Foreground emission from our Galaxy dominates  the COBE data near the Galactic
plane, rendering it unusable for cosmological analyses. We use the galaxy cut
template of \cite{Banday97} to excise pixels with significant Galactic
emission. The cut sky represents an additional challenge for standard
maximum-likelihood analyses. In the absence of this cut, the data sets
represent full-sky coverage and can be decomposed in terms of orthogonal
spherical harmonics.  The resulting coefficients yield the power spectrum of
the CMB and hence the spectral index $n$. Once the galaxy cut is imposed, the
spherical harmonic functions are no longer orthogonal on the remaining pixels.
Any attempt to to obtain a harmonic expansion will result in the aliasing of
power between modes and an inaccurate power spectrum. Though a new orthogonal
set of basis functions can be computed for the cut sky (G\'{o}rski, 1994), this
is an $N^3$ problem as well.  Since neural networks estimate cosmological
parameters  using the real-space pixel values, they need not take the detour
through the power spectrum, and do not suffer aliasing of power. We simply
impose the cut for galactic emission and train each network using only the
remaining high-latitude pixels. As the network is trained ,
it automatically learns the effect of cuts in
the data, without requiring any symmetries in the cut data  
(see {\it e.g.} Oh \etal\ 1999).

The final step in processing our raw theory maps is to fit and remove
the dipole and quadrupole moments. Our peculiar motion, due to the our
motion in the galaxy and the galaxy's motion through the universe, gives
rise to a non-cosmological dipole in the observed CMB anisotropy. 
The quadrupole is removed because it is dominated by local galactic
emission and at the same time containes very little cosmological information.

%---------- Training, Sampling and results ------------
\subsection{Training, sampling and results}

\begin{figure}
\plotone{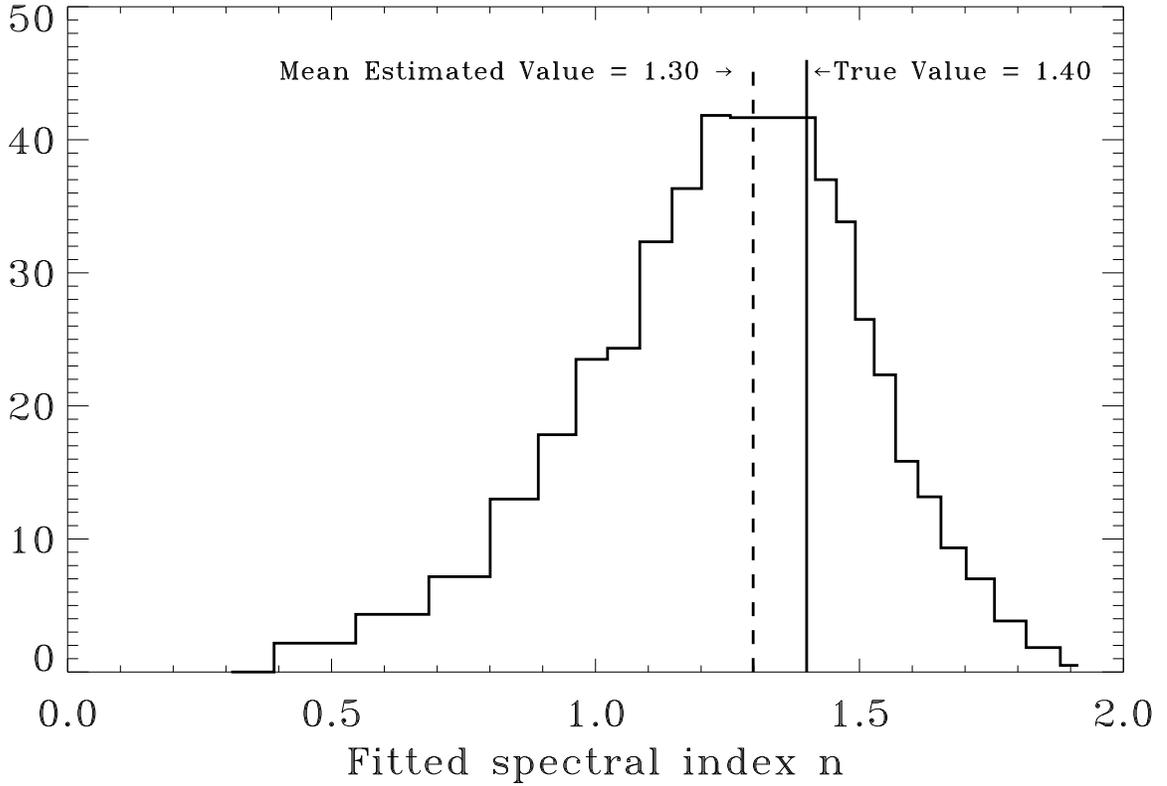}
\caption{
Fitted spectral index $n_{\rm obs}$ 
derived from 1000 realizations of CMB anisotropy sky maps with
$n_{\rm in}=1.25$.
The dotted line is for an initial training
range of $n^{(0)}=0.5$ and $n^{(1)}=1.5$
while the solid line is the distribution for the final
range of $n^{(0)}=0.8$ and $n^{(1)}=1.4$.
The fitted values correctly peak at the input value
(vertical solid line),
despite never having trained on this parameter value.
}
\label{param_hist}
\end{figure}

We generate simulated COBE maps for fixed $Q=20\mu K$ and use the spectral
index $n$ as the parameter to estimate. Working at the COBE-DMR resolution and
after the galaxy cut, this leaves an input
pattern of $\Ninput=3881$ pixels.  We use $N_{\rm Hidden} = 600$ and $N_{\rm
Train}=12000$, along with the learning rate $\eta= 0.0188$ and
initialize the weights with a zero mean Gaussian of width $0.35$
(determined to be the optimal choices).
We globally rescale the input patterns so the variance for {\em all} patterns
in both training sets is unity. 
We can not do this per training set since the variance of
a sky map is dependent on the spectral index $n$.
We train 50 networks over
the  parameter range $n^{(0)}=0.0$ and $n^{(1)}=2.0$.
To build the probability distribution Eqn \ref{eq-prob}, we use
$K=17$ parameter values ranging from $n^{(0)}$ and $n^{(1)}$, with
$N=1000$ sample patterns per sample set.
Figure \ref{param_hist} plots the probability distribution of $n_{\rm obs}$ for a set
of 1000 samples of $n_{\rm in}=1.40$ (dotted line). 
This  distribution matches with separation of the training
sets and we need this only training iteration.
We recover $\bar n_{\rm obs}=1.30$, with 68\% of the samples between
$1.07$ and $1.51$.
In terms of our Bayesian analysis, for $n\sim 1$, we determine
 $\sigma = 0.35$, in agreement with the traditional maximum
likelihood analysis of the COBE-DMR 2 year data \cite{Gorski94b}.

Note that the neural network algorithm recovered the correct spectral index
even though none of the networks used were trained at this value. The
uncertainty, derived from the width of the probability distribution of 
$n_{\rm obs}$, is comparable to the value predicted by 
the maximum likelihood method.
Neural networks can recover cosmological parameters from CMB data sets with
comparable precision as maximum likelihood techniques, but using $N^{1.5}$
calculations instead of $N^3$.

%===================================================================%
%                     Section: CPU Scaling                          %
%===================================================================%
\section{Scaling of Computational Cost}

Neural nets have many desirable characteristics for parameter estimation with
mega-pixel CMB maps. They operate globally on the data and return unbiased
estimates of the underlying parameter values. They automatically account for
data gaps, instrument noise, and other features peculiar to a particular data
set. Most importantly, the computational costs are low enough to allow
extension to the mega-pixel data sets expected in the near future.

The dominant contribution to the CPU cost is the array multiplication 
associated with the weights.  This multiplication is performed twice per
training pass: once for evaluating the pattern, then again for back propagating
the error.  Each operation scales as the number of weights.  Our total
computational cost for training a network is 
\beq
N_{\rm CPU} = 2N_{\rm Train} \left(N_{\rm Hidden}\left(\Ninput+2\right)+1\right)
\eeq
How this cost scales with $\Ninput$ depends on the problem being considered.

%---------- CMB -> per data set: How we model this. ------------
We derive
the scaling for CMB maps by analyzing the CPU costs as progressively larger and
larger areas of the sky are covered. In this way, new information is introduced
into the data sets as the patch size increases. The S/N ratio per pixel is
fixed in this scheme, reflecting the trend in current experiments of scanning
ever larger portions of the sky at (roughly) constant S/N ratio per pixel.

\begin{figure}
\plotone{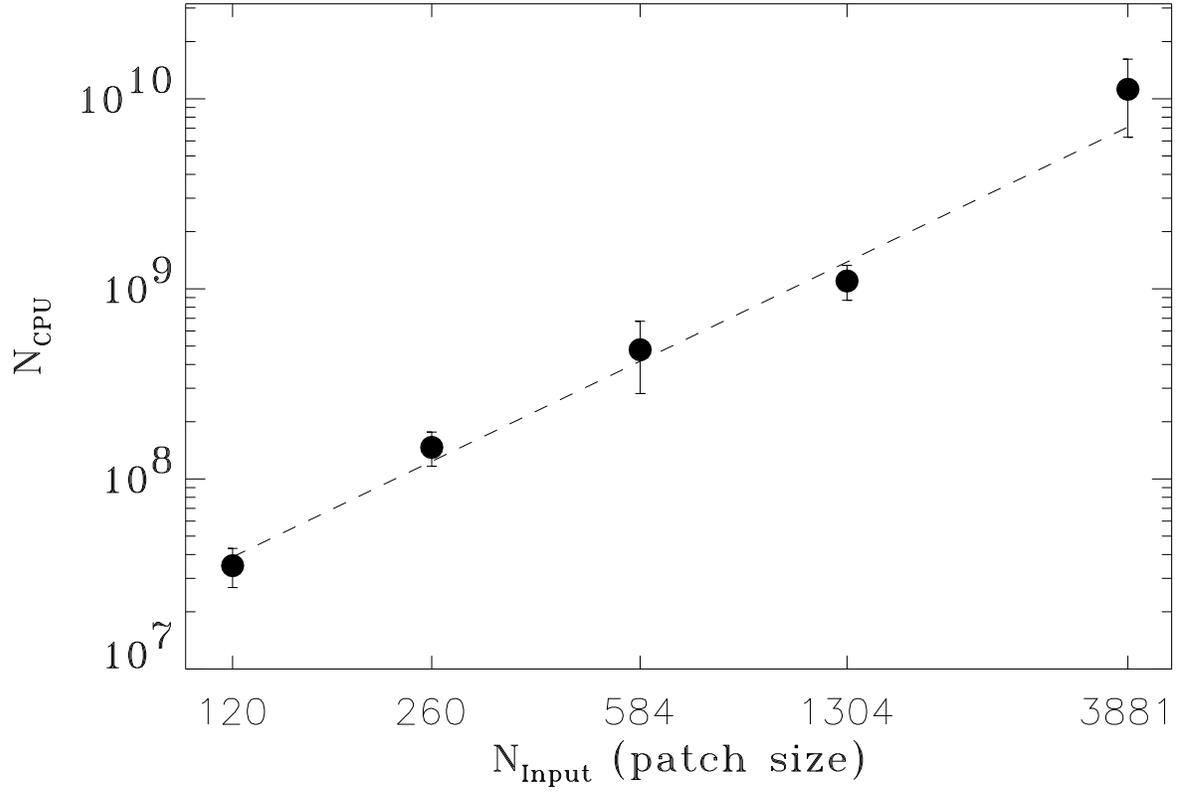}
\caption{
 Scaling of computational costs for CMB anisotropy. Working at a fixed
sky map resolution, we vary the patch size that is examined. This holds
the S/N per pixel fixed, but new information is introduced as the patch
size increases. The solid line represents a power law fit of
$N_{\rm CPU} \sim \Ninput^{1.5}$.
}
\label{ncpu}
\end{figure}

We select circular patches of sky centered at the north zenith.  The range of
patch sizes are chosen to cover 1.5 orders of magnitude in $\Ninput$. 
The number of training passes and the number of hidden units depends on
the number of input pixels, {\it i.e.}, the patch size.
For each patch size, we determine 
which  $N_{\rm Hidden}$ gives the best $N_{\rm CPU}$ for a fixed training accuracy.
For a range of $N_{\rm Hidden}$, we
train multiple networks  on large sets of patterns,
also varying the learning rate and width of the Gaussain used to
initialize the weights.
As the networks are being
trained, we monitor their ability  to correctly classify independent samples
via $\FC$ introduced earlier.
Once this ability passes a pre-set threshold,  we know $N_{\rm Train}$ for each
network, and hence $N_{\rm CPU}$.  We repeat for multiple networks to estimate
the uncertainty in the CPU cost.
For any fixed patch size, we find the optimal learning rate
strongly depends on $N_{\rm Hidden}$, while
the results for $N_{\rm CPU}$ are not strongly  dependent
on the precise value of $N_{\rm Hidden}$.  Above a minimum value, as $N_{\rm
Hidden}$ increases,  the number of training passes needed decreases in such a
way that  $N_{\rm CPU}$ remains constant over a wide range of $N_{\rm
Hidden}$.  
The results
are shown in Figure \ref{ncpu}. The computational cost for training a CMB
network scales according to $N_{\rm CPU}\sim \Ninput^{1.5}$, a
considerable improvement over the $\Ninput^3$ scaling behavior for a
maximum likelihood analysis.

%===================================================================%
%                     Section: Discussion                           %
%===================================================================%
\section{Discussion}
%---------- Neural Nets as viable complement to current methods ------------

We have shown neural networks can be used as a tool for astrophysical
parameter estimation. For specificity, we have worked in the cosmological context
of CMB anisotropy maps where the stochastic nature of the problem is fundamental. 
The results are insensitive  to noise levels and sampling schemes typical of large
astrophysical data sets and provide parameter estimation
comparable to maximum likelihood techniques. 

If we classify parameter estimation techniques as to whether they
are forward or reverse algorithms, we see the real strength of neural
networks.
Maximum-likelihood methods are an
example of reverse algorithms.
They start with the statistic under consideration
and work backwards, inverting a covariance matrix, 
to the likelihood function
used to compare different parameter choices.
Forward algorithms provide a way to avoid the high computational costs of
inverse methods. Typically, it is much simpler to generate model predictions 
at each sampled point in parameter space  than to compute the matrix inverse
and determinant
required for  maximum likelihood techniques. Forward algorithms trade many
realizations of synthetic data sets computed at specific parameter values for
the computationally infeasible matrix inversion. Neural networks are such an
algorithm; synthetic data sets are used to both train and sample the
networks. This gives us our speed improvement.

\begin{figure}
\plotone{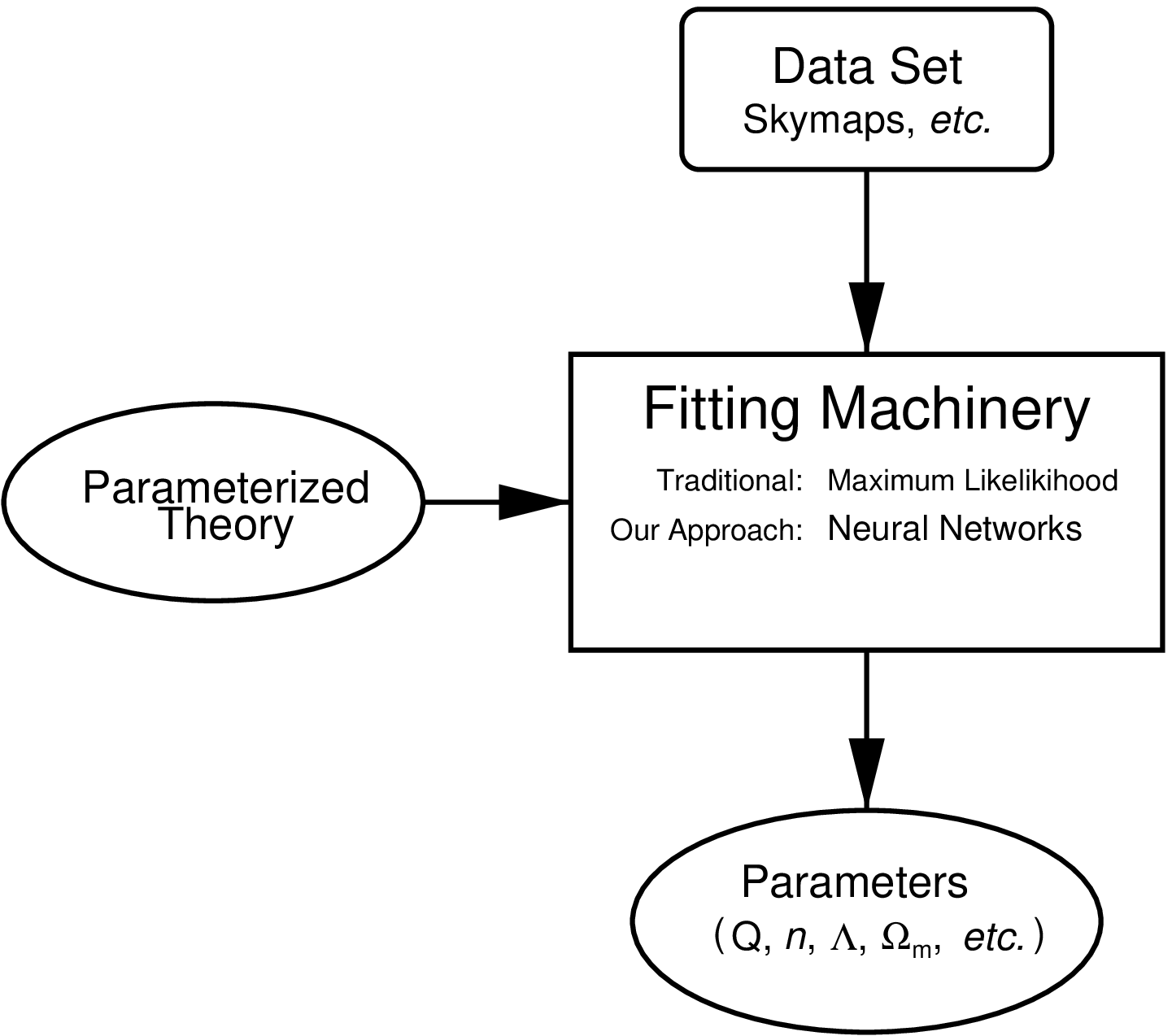}
\caption{
Our basic model for parameter estimation from a data set. We have to {\it a
priori} assume a model to compare the observed data set against; what we pay
attention to is the machinery for performing this comparison. Maximum
likelihood methods, the {\it de facto} standard in the CMB community must
assume a model, just as must be done with currently proposed neural network
method.
}
\label{fitting-model}
\end{figure}

Since either maximum likelihood or neural networks can be viewed as the
``machinery'' for parameter estimation, the fundamental information
flow stays the same (see Figure \ref{fitting-model}). The statistical
confidence levels for the fitted parameters are always accessible. When
the ``machinery'' is sampled with independent  synthetic data, we can
determine the probabilities for making correct or incorrect parameter
identifications. Such sampling also gives us direct access to the
statistical power \cite{PhillipsKogut01}. While training, the information
distinguishing the different parameters is encoded in the weights.
Interperting the resulting weight matrices is not usually possible
(as compared to the Fisher matrix, Eqn \ref{fisher_def}). Using 
independent sampling
of the network to derive the probability distributions needed for,
{\it e.g.} Bayesian analysis, means we do not need direct access to the
information in the weight matrices.

%---------- Situations where NNets might be best method: ------------
Neural networks do not require that we
specify one single statistic of {\it a priori} interest,
a limitation of maximum likelihood
As the network is trained, it determines how it will discriminate.
The information required to separate different parameter points
comes from the training set simulations.
This lack of a need for a goodness-of-fit function can make
neural networks ideal for questions that have been traditionally
hard to to answer. These include probing the global topology of the
universe 
\cite{Lachieze-Rey95,Levin98}
and exploring the viable of non-Gaussian models of
structure formation. Both of these problems involve detecting global
features in the data sets and there is no strong consensus as to the
best statistic to use. The often used power spectrum fails to
capture enough information to decisively test either hypothesis.
If there is a non-trivial topology to the universe, then the
isotropy of the universe assumed when using the power spectrum is
broken. Neural networks may turn out to be the ideal method for
detecting the global features that would be present if indeed the
universe does have a non-trivial topology. Any non-Gaussianity
present in the primordial fluctuations that seed structure formation
is beyond the power spectrum's ability to identify, since it measures
the second moment. The bispectrum 
\cite{Heavens98,Ferreira98,PhillipsKogut01}
(the harmonic conjugate of the
three point correlation function) is often used to detect evidence
of a departure from pure Gaussian. Neural networks, when combined
with a non-Gaussian theory, will provide a complement to bispectrum
tests.

%---------- Other problems in astrophysics: ------------
Along with the explicit example we have presented here of Cosmic Microwave
Background anisotropy data, neural networks as a tool for parameter
estimation has application to other types of data. 
Redshift surveys such as the 2-Degree Field and
the Sloan Digital Sky Survey measure the redshift and position on the sky of a
large number of galaxies ($N \sim 10^6$), sampling the quasi-linear regime
$~\sim 100 h^{-1}$ Mpc where $h$ is the Hubble constant in units 
100 km s${}^{-1}$ Mpc${}^{-1}$.
The observed redshift is the sum of the Hubble flow and the peculiar velocity
induced by gravitational acceleration in the evolving density field. Coherent
flows on large scales  produce artifacts in the redshift distribution compared
to real space. Galaxies on the far side of an overdensity tend to flow toward
the center (hence toward the observer) so that their peculiar velocities
subtract from the Hubble flow, making them appear closer than they really are.
Galaxies on the near side move the opposite direction, so their peculiar
velocities add to the Hubble flow. The net result is an apparent enhancement 
in the galaxy density in redshift space on scales of superclusters, compressing
the region along the line of sight to the observer. The amplitude of this
``bull's-eye'' effect  depends on the matter density $\Omega_m$ on scales
comparable to superclusters of galaxies and can be used to determine $\Omega_m$
in model-independent fashion \cite{Praton97,Melott98}.

Estimating $\Omega_m$ from distortions in redshift space has several problems
in practice. The first is defining a statistic to quantify the bull's-eye
enhancement in concentric rings about the origin. \cite{Melott98} use a large
number of simulations to develop an empirical statistic defined as the ratio of
rms spacing between upcrossings in isodensity contours in the redshift (radial)
direction  to that in the orthogonal (azimuthal) direction. It is thus a local
statistic in that it compares high-density regions only to other nearby
regions, and operates only on a single slice of redshift space after smoothing
and contouring.

Neural nets, by contrast, offer a {\it global} test by comparing each region of
the density field to all other regions simultaneously, and can easily be
extended across the entire three-dimensional survey. No {\it a priori}
statistic need be identified, nor do neural nets require contouring of the
density field, thus avoiding the need to ``fine-tune' the selection of contour
levels. 

Neural networks offer a promising approach to
cosmological parameter estimation, 
where the statistical properties of the primordial matter and energy
distribution provide one of the few falsifiable tests  of the standard
inflationary paradigm. They do this at a computational cost much lower
than traditional maximum likelihood methods.
In the context of CMB anisotropy maps, this cost, $O(\Ninput^{1.5})$,
is better or comparable to the best approximate methods.

%-----------------------------------------------------------------------

%-----------------------------------------------------------------
%-------------------------------------------------------------------------
%-------------------------------------------------------------------------
% --------- Figure 1: fitting-model --------------------------

% --------- Figure 2: Sample Neural net Output  --------- 
%--------------------------------------------------------------------

% --------- Figure 3: Neural net CMB parameter histogram --------- 

% ------------------- Figure 4: CPU cost for sine wave ------------------- 
%-----------------------------------------------------------------------------
%-----------------------------------------------------------------------------

\end{document}